\documentclass[11pt]{article}
\pdfoutput=1
\usepackage{jheppub}
\usepackage{amsmath}
\usepackage{bm}
\usepackage{slashed}
\usepackage{graphicx}

\newlength{\dummysp}
\newcommand{\beq}{\begin{eqnarray}}
\newcommand{\eeq}{\end{eqnarray}}

\newcommand{\gappeq}{\mathrel{\rlap {\raise.5ex\hbox{$>$}}
{\lower.5ex\hbox{$\sim$}}}}
\newcommand{\lappeq}{\mathrel{\rlap{\raise.5ex\hbox{$<$}}
{\lower.5ex\hbox{$\sim$}}}}

\newcommand{\ben}{\begin{enumerate}}
\newcommand{\een}{\end{enumerate}}

\newcommand{\bit}{\begin{itemize}}
\newcommand{\eit}{\end{itemize}}

\def\[{\left [}
\def\]{\right ]}
\def\({\left (}
\def\){\right )}

\def\S{{\mathbb S}}
\def\Z{{\mathbb Z}}

\title{Generalized 't Hooft anomalies on non-spin manifolds}
   
 \author[a]{Mohamed M. Anber,}\author[b]{Erich Poppitz} 
  
\affiliation[a]{Department of Physics, Lewis $\&$ Clark College, Portland, OR 97219, USA}
\affiliation[b]{Department of Physics,   University of Toronto, 
Toronto, ON M5S 1A7, Canada}
\emailAdd{manber@lclark.edu}\emailAdd{poppitz@physics.utoronto.ca}    

\abstract{We study the mixed anomaly between the discrete chiral symmetry and general baryon-color-flavor (BCF) backgrounds in $SU(N_c)$ gauge theories with $N_f$ flavors of Dirac fermions in representations ${\cal R}_c$ of $N$-ality $n_c$, formulated on non-spin manifolds. We show how to study these theories  on $\mathbb{CP}^2$ by turning on general BCF fluxes consistent with the fermion transition functions. We consider several examples in detail  and argue that  matching the anomaly on non-spin manifolds places stronger constraints on the infrared physics, compared to the ones on spin manifolds (e.g.~$\mathbb{T}^4$).   We also show how to consistently formulate various chiral gauge theories on non-spin manifolds.  }

\begin{document}

\maketitle

\flushbottom

\section{Introduction}
Anomaly matching conditions provide a rare exact  constraint on the infrared (IR) behavior of strongly coupled gauge theories \cite{tHooft:1979rat}. To study the matching of anomalies, one probes the theory with nondynamical (background) gauge fields for its anomaly-free global symmetries. Any violation of the background gauge invariance due to the resulting 't Hooft anomalies should exactly match between the ultraviolet (UV), usually free, and IR descriptions of the theory. In the past, these consistency conditions have been applied to ``0-form" symmetries, acting on local fields. For example, anomaly matching was instrumental in the study of models of quark and lepton compositeness in the 1980s (see the review \cite{Rosner:1998wh}) or of Seiberg duality in the 1990s  \cite{Seiberg:1994pq}. 

Recently, it was realized that the scope of anomaly matching is significantly wider than originally thought \cite{Gaiotto:2014kfa,Gaiotto:2017yup,Gaiotto:2017tne}. Turning on general background fields---corresponding to global, spacetime, continuous, discrete, $0$-form, or higher-form symmetries, consistent with their faithful action---was argued to lead to new UV-IR anomaly matching conditions. We refer to them as ``generalized 't Hooft anomalies." 
The study of these generalized anomalies is a currently active area of research with contributions coming from the high-energy, condensed matter, and mathematical  communities. We do not claim to be in command of all  points of view and only give a list of references written from a (largely) high-energy physics perspective and pertaining to theories somewhat similar to the ones discussed in this paper \cite{Komargodski:2017smk,Sulejmanpasic:2018upi,Tanizaki:2018wtg,Tanizaki:2018xto,Wan:2018bns, Cordova:2019jnf, Cordova:2019uob,Anber:2019nfu,Bolognesi:2019fej,Cordova:2019bsd,Wan:2019soo,Wan:2019oax,Cordova:2019jqi}.

{\flushleft{\bf Summary: }} We continue our study \cite{Anber:2019nze} of the generalized 't Hooft anomalies  in $SU(N_c)$ gauge theories with $N_f$ flavors of Dirac fermions in representations ${\cal{R}}_c$ of $N_c$-ality $n_c$. These theories have exact global discrete chiral symmetries. Considering these theories on $\mathbb T^4$ and turning on the most general 't Hooft flux \cite{tHooft:1979rtg} backgrounds for the global symmetries, consistent with their faithful action, we found  a mixed anomaly between the discrete chiral symmetry and the $U(N_f)/\Z_{N_c}$ baryon-color-flavor, or ``BCF", background. We showed that matching this BCF anomaly  imposes new constraints on possible scenarios for IR physics, in addition to those imposed by the ``traditional" 0-form 't Hooft anomalies. When these theories are coupled to axions, the axion theory is also constrained by anomalies  \cite{Anber:2020xfk}.

In this paper, we consider the fate of the BCF anomalies in the same class of theories, but now formulated on non-spin manifolds. We are motivated by the study of QCD(adj)  \cite{Cordova:2018acb}, which   showed that 't Hooft anomalies in theories with fermions on  non-spin backgrounds  impose additional constraints.
It is known that manifolds that do not permit a spin structure \cite{Geroch:1968zm,Geroch:1970uv} can accommodate theories with fermions, but  only   if appropriate gauge fluxes are turned on   \cite{Hawking:1977ab}. These  fluxes can correspond to dynamical or background fields, as in the recent studies   \cite{Cordova:2018acb,Wang:2018qoy,Bi:2018xvr,Davighi:2020bvi}. We focus on the canonical example of non-spin manifold, $\mathbb{CP}^2$. It has the advantage of allowing for an explicit (and pedestrian\footnote{See Appendix \ref{CP2 space} for details of    $\mathbb{CP}^2$ and Appendix \ref{gauge fields and fermions} for an explicit description of how to consistently turn on 't Hooft fluxes on $\mathbb{CP}^2$ in theories with fermions in general representations. This discussion complements the more abstract mathematical descriptions existing in the literature.  At the end, the anomaly depends only on topological information. However, considering explicit gauge and gravity backgrounds ('t Hooft fluxes in a $\mathbb{CP}^2$ background) provides a more ``pedestrian" route to see the anomaly, which might be more familiar for many physicists.}) discussion of the salient points. We describe in detail how to turn on background $U(N_f)/\Z_{N_c}$ fluxes on $\mathbb{CP}^2$  and derive the resulting BCF anomaly on non-spin backgrounds. 
The final result of our analysis is that the BCF anomaly matching conditions on $\mathbb{CP}^2$ are equal or stronger than  those obtained on $\mathbb{T}^4$. We use several examples to show that the BCF anomaly on $\mathbb{CP}^2$   further constrains various scenarios for the IR dynamics. 

{\flushleft{\bf Organization of this paper:}} In Section \ref{vectorlikedef}, we define the class of theories we study. In Section \ref{thooftcp2}, inviting the reader  to also consult  Appendices \ref{CP2 space}  and \ref{gauge fields and fermions}, we explain how to turn on 't Hooft fluxes on $\mathbb{CP}^2$ for the baryon, color, and flavor gauge fields, consistent with the faithful action of the global symmetries in the representation ${\cal{R}}_c$.

 In Section \ref{chiralsection}, we temporarily divert to show how to put chiral gauge theories in non-spin backgrounds; however, we leave their study for the future.

In Section \ref{bcfanomalies}, we study the mixed 't Hooft anomalies of the discrete chiral symmetry with the BCF fluxes on $\mathbb{CP}^2$, discuss the conditions imposed on the IR spectrum, and compare with the case of $\mathbb{T}^4$ studied previously.

In Section \ref{examples}, we present several examples. In Section \ref{qcdadj} we discuss QCD(adj). Our intention is to use the present study to investigate the  various scenarios for IR behavior, whose consistency has been recently elaborated upon in \cite{Anber:2018tcj,Cordova:2018acb,Bi:2018xvr,Wan:2018djl,Poppitz:2019fnp,Cordova:2019bsd,Cordova:2019jqi}. In Section \ref{su6}, we study an $SU(6)$ gauge theory with a single Dirac flavor in the two-index antisymmetric representation and, in Section \ref{su4k2}, its generalization to $SU(4k+2)$ with a single flavor of  two-index symmetric or antisymmetric representations. In both cases, we argue that scenarios for IR physics consistent with the 0-form 't Hooft anomalies are further constrained by studying them on $\mathbb{CP}^2$. In particular, we focus on exotic phases\footnote{The examples of Sections \ref{su6} and \ref{su4k2} were also  studied in refs.~\cite{Cordova:2019bsd,Cordova:2019jqi}, which  argued that   an IR gapped phase with unbroken global symmetries cannot occur. }
 with massless composite fermions, and argue that the TQFT which must accompany the massless composites has to reproduce a more restrictive anomaly on $\mathbb{CP}^2$.
 
Appendices  \ref{CP2 space} and \ref{gauge fields and fermions} contain many relevant formulae regarding $\mathbb{CP}^2$ and fermions. At the end of Appendix \ref{gauge fields and fermions}, we find several classes of theories which can be formulated on $\mathbb{CP}^2$ by turning on of only dynamical gauge backgrounds, i.e.~by only modifying the gauge bundles summed over. These gauge theories share a feature common with examples discussed in  \cite{Bi:2018xvr,Wang:2018qoy}: they have only bosonic gauge invariant operators and can be taught as emergent descriptions near quantum critical points of purely bosonic systems.

\section{Baryon-Color-Flavor (BCF) 't Hooft fluxes on $\mathbb{CP}^2$ for vector-like theories}

In this Section, we describe in great detail (in conjunction with Appendices \ref{CP2 space}  and \ref{gauge fields and fermions}) how to introduce background fluxes in the baryon-number, color, and flavor directions on $\mathbb{CP}^2$. We carry out our construction for vector-like theories. However, this setup can be easily adapted for chiral theories (such as the Standard Model), as we show at the end of this Section. 

\subsection{Vector-like theories}
\label{vectorlikedef}

We consider  $SU(N_c)$ gauge theories with $N_f$ flavors of Dirac fermions  transforming in a representation ${\cal R}^c$  of N-ality $n_c$.\footnote{The $N$-ality of a representation ${\cal R}$ of $SU(N)$ is   the number of boxes of the Young tableau of ${\cal R}$ modulo  $N$.}  The gauge group that faithfully acts on the fermions is $\frac{SU(N_c)}{\mathbb Z_p}$, where $p=\mbox{gcd}(N_c,n_c)$; thus, the fermions are charged under a $\mathbb Z_{\frac{N_c}{p}}$ subgroup of the center of $SU(N_c)$. After modding out the redundant symmetries, we find that the $0$-form global symmetry of the theory is
\begin{eqnarray}\label{globalG}
G^{\scriptsize\mbox{global}}=\frac{SU(N_f)_L\times SU(N_f)_R\times U(1)_B\times \mathbb Z_{2\,\mbox{dim}({\cal R}^f)T_{{\cal R}^c}}}{\mathbb Z_{\frac{N_c}{p}}\times\mathbb Z_{N_f}\times \mathbb Z_2}\,,
\end{eqnarray}
where $T_{{\cal R}}$ is the Dynkin index of the representation ${\cal R}$ (normalized so that $T_{\Box}=1$) and $\mbox{dim}(\cal R)$ is its dimension. Here, we assume that $\mathbb Z_{2\,\mbox{dim}({\cal R}^f)T_{{\cal R}^c}}$ is a genuine symmetry of the theory; thus, it cannot be absorbed in the continuous part of $G^{\scriptsize\mbox{global}}$ (this can be checked on a case by case basis). $\Z_2$  above denotes fermion number and $\Z_{N_c\over p}$ is in the center of $SU(N_c)$. 

 In addition,  the theory has a $1$-form center symmetry $\mathbb Z_p^{(1)}$ that acts on non-contractible Wilson loops, provided that $\mbox{gcd}(N_c,n_c)=p>1$. Notice that the ultraviolet fermions are taken to transform in the defining representation of the flavor group $SU(N_f)$, and hence, we should use $n_f=1$. Nevertheless, we keep the $N$-ality of the fermions under $SU(N_f)$ an arbitrary integer for the sake of generality. 

\subsection{Generalized 't Hooft fluxes on $\mathbb{CP}^2$}

\label{thooftcp2}

Next, we turn on 't Hooft fluxes (twists) in the baryon-number, color, and flavor directions,  which are compatible with  $\mathbb{CP}^2$ and at the same time lead to consistent transition functions. See Appendix \ref{CP2 space} for a collection of relevant formulae for $\mathbb{CP}^2$.

We first address the compatibility condition.   As we point out in Appendix \ref{gauge fields and fermions},  background gauge fields (both abelian and nonabelian) on $\mathbb{CP}^2$ need to be (anti)self-dual, otherwise they will have  a nonvanishing energy-momentum, and hence,   backreact on the manifold. In order to achieve the (anti)self-duality, we  take the gauge fields to be proportional to the K{\"a}hler $2$-form $K$ of $\mathbb{CP}^2$, eqs.~(\ref{kahler1}, \ref{2-form in polar}):
\begin{eqnarray}
T^aF^a\sim T^aC^aK\,,
\label{gauge fields}
\end{eqnarray}
where $T^a$ stands for the color, flavor, or baryon-number generators, and $C^a$ are constants that will be determined momentarily. 

Second, we come to the problem of defining a consistent gauge theory with matter fields on a manifold ${\cal M}$. Let $G$ be a direct product of semi-simple Lie groups and  $\Psi$ a fermionic matter field transforming under specific representations of $G$. A quantum field theory of $\Psi$ is described in terms of a collection of covers $\{U_i\}$ of ${\cal M}$ (in $\{U_i\}$, $\Psi$  is denoted  $\Psi_i$), along with transition functions $g_{ij}\in G$, defined on the overlap $U_i\cap U_j$ and relating $\Psi_i$ to $\Psi_j$
\begin{eqnarray}
\Psi_i={\cal G}_{ij}\Psi_j\,,
\end{eqnarray}
where
\begin{eqnarray}\label{transition1}
{\cal G}_{ij}=g_{ij}^{B}g_{ij}^{{\cal R}^c}g_{ij}^{{\cal R}^f}g_{ij}^{L}\,,
\end{eqnarray}
such that $g_{ij}^{B, {\cal R}^c, {\cal R}^f}$ are the transition functions of the baryon, color, and flavor groups, while $g_{ij}^L$ is the transition function associated to the spacetime Lorentz group. The matter field in general will transform under representation ${\cal R}^c$ of the color group and representation ${\cal R}^f$ of the flavor group. However, only the N-ality of the representations will matter in what follows.  Consistency requires that the transition functions   satisfy the cocycle conditions on the triple overlap $U_i\cap U_j\cap U_k$:
\begin{eqnarray}
{\cal G}_{ij}{\cal G}_{jk}{\cal G}_{ki}=1\,.
\label{cocycle 1}
\end{eqnarray}
The above cocycle condition does not necessary imply that the strong conditions $g_{ij}^ag_{jk}^ag_{ki}^a=1$ should be met for each of the transition functions in (\ref{transition1}), where $a$ refers to the baryon-number, flavor, color, or Lorentz groups. 

Let $g_{ij}^c$ and $g_{ij}^f$ be the transition functions in the defining representations of the color and flavor groups, respectively.  One, then,  may relax the condition (\ref{cocycle 1}) to the following set of conditions
\begin{eqnarray}
\nonumber
&&g_{ij}^cg_{jk}^cg_{ki}^c=e^{i\frac{2\pi }{N_c}n_{ijk}^{(c)}}\,,\quad g_{ij}^fg_{jk}^fg_{ki}^f=e^{i\frac{2\pi }{N_f}n_{ijk}^{(f)}}\,,\\
&& g_{ij}^{B}g_{jk}^{B}g_{ki}^{B}=e^{-i\pi-i n_c\frac{2\pi}{N_c}n_{ijk}^{(c)}-i n_f\frac{2\pi}{N_f}n_{ijk}^{(f)}}\,,
\label{U1 cocycle}
\end{eqnarray}
on the triple overlap. In this expression $n_c$ ($n_f$) is the color (flavor) N-ality, $n_{ijk}^{(c)}$ ($n_{ijk}^{(f)}$) are integers modulo $N_c$ ($N_f$), while the factor $e^{-i\pi}$ that appears in the last cocycle condition cancels the minus sign   arising from parallel transporting the spinor fields around appropriate closed paths in $\mathbb{CP}^2$, see Appendix \ref{gauge fields and fermions} and \cite{Geroch:1968zm,Geroch:1970uv,Hawking:1977ab}.

 Thus, the $U(1)_B$ bundle provides the flux that is necessary to render the fermions well-defined on the non-spin manifold. 
As a side remark, we note that  this is by no means is the unique choice to put spinors on $\mathbb{CP}^2$: one could also use the fluxes in the color (or flavor) directions to perform the same job. Examples of using only gauge backgrounds (i.e.~modifying only the  gauge bundles being summed over in the path integral) are known in the literature \cite{Bi:2018xvr,Wang:2018qoy,Davighi:2020bvi} and we give a few more at the end of Appendix \ref{gauge fields and fermions}; a common feature of gauge theories where this can be done is their  possible interpretation  as  emergent descriptions near quantum critical points in theories of only bosons \cite{Bi:2018xvr}.

The consistency conditions (\ref{cocycle 1}) or (\ref{U1 cocycle}) guarantee that the Dirac  index  will always be an integer. Since the Dirac  index counts the number of the fermion zero modes in a given gauge/gravity background, the integrality of the index is a necessary condition for the consistency of a given  theory in the background of baryon-color-flavor  't Hooft fluxes in  $\mathbb{CP}^2$. The integrality of the index will be manifest in all the examples we discuss in this paper. 

Having all the ingredients necessary to turn on compatible fluxes on non-spin manifolds, we now choose the color and flavor fluxes in the Cartan directions of the respective groups. Using (\ref{gauge fields}) we write:
\begin{eqnarray}
\nonumber
T^{a(c)}F^{a(c)}&=&\bm H^c\cdot \bm \nu^c m^c K\,,\\
\nonumber
T^{a(f)}F^{a(f)}&=&\bm H^f\cdot \bm \nu^f m^f K\,,\\
F^{B}&=&\left(\frac{1}{2}+\frac{n^c}{N_c}m^c+\frac{n^f}{N_f}m^f\right)K\,.
\label{fluxes in Cartan}
\end{eqnarray}
 Here $\bm H^{c/f}$ are the fundamental representation Cartan generators of $SU(N_{c/f})$, obeying $\mbox{tr}\left[H^a H^b\right]=\delta^{ab}$, and $\bm \nu$ are the weights of the corresponding defining representation, $\bm \nu^a\cdot \bm\nu^b=\delta^{ab}-\frac{1}{N}$ (where $N$ stands for $N_c$ or $N_f$). The fluxes (\ref{fluxes in Cartan}), with integer $m^c$ and $m^f$, are compatible with the cocycle conditions (\ref{U1 cocycle}), see  (\ref{sphere2}), and  the Dirac index is integer in their background. 
The topological charges  are given by
\begin{eqnarray}
Q=\frac{1}{8\pi^2}\int\mbox{tr}\left[ F\wedge F\right]\,.
\label{top charge}
\end{eqnarray}
Then, substituting (\ref{fluxes in Cartan}) into (\ref{top charge}) and using  $\int_{\mathbb{CP}^2}{ K\wedge K \over 8 \pi^2}=\frac{1}{2}$,  we find:
\begin{eqnarray}
\nonumber
Q^c&=&\frac{\left(m^c\right)^2}{2}\left(1-\frac{1}{N_c}\right)\,, \quad Q^f=\frac{\left(m^f\right)^2}{2}\left(1-\frac{1}{N_f}\right)\,,\\
Q^{B}&=&\frac{1}{2}\left(\frac{1}{2}+\frac{n^c}{N_c}m^c+\frac{n^f}{N_f}m^f\right)^2\,.
\label{fractional top charges}
\end{eqnarray}
Adding to this list the gravitational topological charge of $\mathbb{CP}^2$ 
\begin{eqnarray}
Q^G=\frac{1}{192\pi^2}\int \mbox{tr}\left[R\wedge R\right]=-\frac{1}{8}\,,
\label{grav charge}
\end{eqnarray}
we finally obtain the Dirac  index:
\begin{eqnarray}
{\cal J}_D=T_{{\cal R}^c}\mbox{dim}_{{\cal R}^f}Q^c+ T_{{\cal R}^f}\mbox{dim}_{{\cal R}^c}Q^f+\mbox{dim}_{{\cal R}^f}\mbox{dim}_{{\cal R}^c}\left(Q^B+Q^G\right)\,,
\label{Dirac index}
\end{eqnarray}
which is an integer for all the examples we consider below. 

Before moving to examples, it is instructive to compare and contrast the above results with the BCF fluxes on the four-torus $\mathbb T^4$ that we considered before  \cite{Anber:2019nze}. $\mathbb{CP}^2$ has one two-cycle $\mathbb{CP}^1$, and hence, we were able to turn on fluxes along this single cycle (the color and flavor fluxes are labeled by $m^{c,f}$ in (\ref{fractional top charges})). In contrast,  $\mathbb T^4$ has six two-cycles (it suffices to turn on fluxes in the $1$-$2$ or $3$-$4$ planes, respectively, hence we have two integers $m_{12}$ and $m_{34}$ that label the fluxes).\footnote{\label{t4thooft}For the sake of completeness, we give $Q^{c,f,B}$ on $\mathbb T^4$  \cite{Anber:2019nze}:
\begin{eqnarray}
\nonumber
Q^c&=&m_{12}^c m_{34}^c\left(1-\frac{1}{N_c}\right)\,,\quad Q^f=m_{12}^f m_{34}^f\left(1-\frac{1}{N_f}\right)\,,\\
Q^B&=&\left(n_c\frac{m_{12}^c}{N_c}+n_f\frac{m_{12}^f}{N_f}\right)\left(n_c\frac{m_{34}^c}{N_c}+n_f\frac{m_{34}^f}{N_f}\right)\,.
\label{topological charges on a torus}
\end{eqnarray}
}  Since there are more ways to turn on fluxes on $\mathbb T^4$ compared to $\mathbb{CP}^2$,  this may imply that putting the theory on $\mathbb T^4$ can give us more constraining conditions on the IR spectrum. We will see in the next section that this is not true: although $\mathbb{CP}^2$ has only one cycle, it always imposes conditions that are either stronger or at least as strong as the conditions we obtain by putting the theory on $\mathbb T^4$.  
 
\subsection{Comment on chiral theories and the Standard Model with $\nu_R$}
\label{chiralsection}

Here, we slightly divert from our main presentation to note, for the sake of completeness, that by turning on global anomaly-free $U(1)$ fluxes,  chiral gauge theories can also be formulated on non-spin manifolds.

As an example, consider an $SU(5)$ gauge theory with $\mathbf{5^*}$ and $\mathbf{10}$   left-handed Weyl fermions:\footnote{For a discussion of its conjectured IR dynamics, see \cite{Raby:1979my}.}
     $\lambda$ in the anti-fundamental and $\psi$  in the two-index anti-symmetric representations. This theory has an anomaly-free global $U(1)$ that acts on the fermions as $\psi\rightarrow e^{i2\pi \alpha}\psi$ and $\lambda \rightarrow e^{-i2\pi (3 \alpha)}\lambda$. Then, one can easily check that  the flux
\begin{eqnarray}
\nonumber
T^{a(c)}F^{a(c)}&=&\bm H^c\cdot \bm \nu^c m^c K\,,\\
F^{U(1)}&=&-\left(\frac{1}{2}+\frac{2}{5}m^c\right)K\,,
\label{chiral}
\end{eqnarray}
 is consistent with the cocycle condition (\ref{cocycle 1}) for both $\psi$ and $\lambda$. This can be seen by considering the consistency condition  (\ref{sphere2}) on $\mathbb{CP}^2$ for fermions in these two representations, taking into account their different  $U(1)$ charges and $SU(5)$ representations. One can also check the consistency by calculating  the Dirac indices for both $\psi$ and $\lambda$: using $Q^c=\frac{1}{2}(m^c)^2\left(1-\frac{1}{5}\right)$ and $Q^{U(1)}=\frac{1}{2}\left(\frac{1}{2}+\frac{2}{5}  m^c\right)^2$, we obtain
\begin{eqnarray}
\nonumber
{\cal J}_{\psi}&=&T_{\psi}Q^c+\mbox{dim}_{\psi}\left(Q^{U(1)}-\frac{1}{8}\right)=2 m^c (1 + m^c)\,,\\
{\cal J}_{\lambda}&=&T_{\lambda}Q^c+\mbox{dim}_{\lambda}\left((3)^2Q^{U(1)}-\frac{1}{8}\right)=5 + 9 m^c + 4( m^c)^2\,,
\end{eqnarray}
 which are integers. Notice that the total number of upper minus lower $SU(5)$ indices of the zero modes is a multiple of 5 (and the total number of zero modes is even for odd $m_c$), so that a gauge invariant ``'t Hooft vertex" using the zero modes can be written.
 
Let us also mention that the Standard Model can be formulated on a non-spin manifold,  provided that  right-handed neutrinos are added.\footnote{In the absence of right-handed neutrinos one finds that $U(1)_{B-L}$ is broken by gravitational instantons.} In this case one can turn  on a fractional flux in the global $U(1)_{B-L}$ in order to cancel the $e^{i\pi}$ ambiguity  that results from putting the quarks and leptons on $\mathbb{ CP}^2$. By computing the indices, as above, it is easy to see that gauge and Lorentz invariant terms can be constructed out of the zero modes.
The $U(1)_{B-L}$ can further be promoted to a gauge symmetry, broken by a charge-2 Higgs. For related discussions see \cite{Wan:2019fxh,Davighi:2019rcd} as well as the remarks on the $Spin(10)$ grand unified theory in \cite{Wang:2018qoy}.

In the two examples mentioned in this Section, formulating the theory on $\mathbb{CP}^2$ does not lead to new 't Hooft anomalies of the type discussed here, as these theories only have continuous chiral symmetries whose anomalies are matched irrespective of the integrality of the topological charges\footnote{See \cite{Anber:2019nze} for a lucid explanation why continuous chiral symmetry transformations in BCF backgrounds do not impose further constraints.}. Further study of chiral theories is left for the future.

\section{Anomalies in the background of BCF fluxes on $\mathbb{CP}^2$}

\label{bcfanomalies}

We now return back to our main theme and  examine the fate of the axial symmetries of vector-like theories as we put them in the background of BCF fluxes.  In order to reduce notational clutter, we assume that the theory enjoys a genuine discrete $\mathbb Z_{q_g}$ axial global symmetry, which becomes anomalous  in the background of BCF fluxes.  We denote by $D^{c,f,B,G}$ the anomaly coefficients that accompany the color, flavor, baryon-number, and gravitational topological charges. The UV values of these coefficients, $D^c_{UV}, D^f_{UV}, D^B_{UV}, D^G_{UV}$, are equal to twice the pre-factors that multiply $Q^{c,f,B,G}$, respectively, in the Dirac  index (\ref{Dirac index}): these are group-theoretical values and they do not depend on whether we turn on integer or fractional fluxes or whether we put the theory on  spin or non-spin manifolds. To summarize, upon performing a global $\mathbb Z_{q_g}$ axial transformation on the fermions, the UV partition function acquires the phase
\begin{eqnarray}
{\cal Z}_{UV}|_{\mathbb Z_{q_g}}\rightarrow {\cal Z}e^{i \frac{2\pi}{q_g}\left(D^c_{UV}Q^c+D^f_{UV}Q^f+D^B_{UV}Q^B+ D^G_{UV}Q^G\right)} = {\cal Z}e^{i \frac{2\pi}{q_g} 2 {\cal J}_D}~,
\label{UV Z}
\end{eqnarray}
where ${\cal  J}_D$ is the Dirac index (\ref{Dirac index}). This phase is a manifestation of a 't Hooft anomaly between the $0$-form $\mathbb Z_{q_g}$  symmetry and a general BCF background.  

Now, we assume that the $0$-form (``traditional") 't Hooft anomalies, which correspond to integer values of $Q^{c,f,B,G}$, can be matched by a set of fermion composites deep in the IR on a spin manifold. Upon performing a $\mathbb Z_{q_g}$ transformation in the IR, the partition function transforms as
\begin{eqnarray}
{\cal Z}_{IR}|_{\mathbb Z_{q_g}}\rightarrow {\cal Z}e^{i \frac{2\pi}{q_g}\left(D^c_{IR}Q^c+D^f_{IR}Q^f+D^B_{IR}Q^B+ D^G_{IR}Q^G\right)}\,,
\label{IR Z}
\end{eqnarray}
where $D^c_{IR}, D^f_{IR}, D^B_{IR}, D^G_{IR}$ are the anomaly coefficients computed using the IR spectrum of composites.
Since we are matching a discrete anomaly, the coefficients $D^{c,f,B,G}$ need not be exactly matched between the UV and IR. Instead, $D^{c,f,B}$ are matched modulo $q_g$:
\begin{eqnarray}
D^{c,f,B}_{UV}-D^{c,f,B}_{IR}=q_g\ell^{c,f,B}\,,
\label{condition 1}
\end{eqnarray}
for integers $\ell^{c,f,B}$. The coefficients  $D^G$ are matched only modulo $q_g/2$: there is an integer $\ell^G$ such that
\begin{eqnarray}
D^{G}_{UV}-D^{G}_{IR}=\frac{q_g}{2}\ell^{G}\,.
\label{condition 2}
\end{eqnarray}
This is true since the gravitational topological charge of a spin manifold is an even number.\footnote{Notice that $q_g$ is an even number, since $\mathbb Z_{q_g}$ has to contain $\mathbb Z_2$ as a subgroup in any theory that preserves its Lorentz symmetry.}

Now, we would like to check whether the same set of IR composite fermions can also match the BCF anomaly as we turn on fractional fluxes on a non-spin manifold. Before doing that, we first note that if a non-spin manifold admits an elementary spinor $\Psi$, then by virtue of (\ref{cocycle 1}) and (\ref{U1 cocycle}) a composite of these spinors can always be defined. Also, one can easily see the spin-charge relation of the composites: a fermion (boson), made of an odd (even) number of $\Psi$, carries an odd (even) charge under $U(1)_B$. 

Thus, using (\ref{UV Z}),  (\ref{IR Z}), (\ref{condition 1}), and (\ref{condition 2}), we obtain the matching condition:
\begin{eqnarray}
\frac{{\cal Z}_{UV}|_{\mathbb Z_{q_g}}}{{\cal Z}_{IR}|_{\mathbb Z_{q_g}}}=e^{i2\pi\left(\ell^c Q^c+\ell^f Q^f+\ell^B Q^B+\frac{\ell^G}{2} Q^G\right)}=1\,,
\end{eqnarray}
or in other words
\begin{eqnarray}
\ell^c Q^c+\ell^f Q^f+\ell^B Q^B+\frac{\ell^G}{2} Q^G \in \mathbb Z
\label{condition on topological charges}
\end{eqnarray}
for all fractional charges $Q^{c,f,B,G}$ given in (\ref{fractional top charges}) and (\ref{grav charge}). 
The condition (\ref{condition on topological charges}) can be translated into the following set of conditions on $\ell^{c,f,B,G}$, which can be obtained by turning on and off the fluxes in the various directions: 
\begin{eqnarray}
\nonumber
&&(i)\, \ell^c N_c(N_c-1)+\ell^B n_c(n_c+N_c)\in 2 N_c^2\,\mathbb Z\,,\\
\nonumber
&& (ii)\, \ell^f N_f(N_f-1)+\ell^B n_f(n_f+N_f) \in 2 N_f^2\,\mathbb Z\,,\\
\nonumber
&& (iii)\,\ell^c N_f^2N_c(N_c-1)+\ell^f N_c^2N_f(N_f-1)+\ell^B(n_cN_cN_f^2+n_fN_fN_c^2)\\
\nonumber
&&\quad\quad+\ell^B\left(n_cN_f+n_fN_c\right)^2 \in 2 N_c^2 N_f^2\, \mathbb Z\,,\\
&&(iv)\,2\ell^B-\ell^G\in 16\,\mathbb Z\,.
\label{conditions for absence of massless spectrum}
\end{eqnarray}
 The importance of the above conditions is as follows: if no integers $\ell^{c,f,B,G}$ that satisfy (\ref{conditions for absence of massless spectrum}) can be found, then composite fermions cannot solely match the BCF anomaly. Thus, either the composites do not form in the IR, or   they are accompanied by a partial breaking of $\mathbb Z_{q_g}$, due to some higher dimensional fermion condensate that leaves the continuous flavor symmetries intact,  and/or an IR TQFT. 

For example, setting\footnote{Notice that gauge invariant composites   have $\ell^c=1$ in the vectorlike theories we consider:  using $D_{IR}^c = 0$, since the composites are color singlets, we have $\ell^c = {D_{UV}^c\over q_g} = \frac{T_{{\cal R}^c \mbox{dim}_{{\cal R}^f}}}{T_{{\cal R}^c \mbox{dim}_{{\cal R}^f}}} = 1$.} $\ell^c=1$,  it is straightforward to check that no integers $\ell^{c,f,B,G}$ exist that satisfy (\ref{conditions for absence of massless spectrum}) if $N_f\geq 2$ and one of the following two conditions are met:
\begin{eqnarray}
\nonumber
&&(i)\,\mbox{gcd}(N_c,N_f)>1\,,\\
&&(ii)\,\mbox{gcd}(N_c,n_c)>1\,.
\label{no go conditions}
\end{eqnarray}
We call the inequalities (\ref{no go conditions}) the ``no-go condition" on the composites  (we stress that they apply provided that $N_f\geq 2$ and recall that $n_f=1$). In the special case $N_f=1$, one needs to replace  (\ref{no go conditions}) by other sets of conditions that we do not quote here; they can be checked on a case by case basis using the first and last conditions in (\ref{conditions for absence of massless spectrum}).

Now a few comments are in order:
\begin{enumerate}

\item \label{first point} The first three conditions (\ref{conditions for absence of massless spectrum}) are functions of  $\ell^{c,f,B}$, while the fourth condition is a function of two variables only, $\ell^G$ and $\ell^B$. Therefore, if $\ell^{c,f,B}$ can be found to satisfy conditions $(i)$ to $(iii)$, then it is always trivial to find $\ell^G\in \mathbb Z$  that satisfies condition $(iv)$. 

\item Given \ref{first point} above, one expects that turning on gravitational background does not alter the conditions that are needed to find a set of composites in the IR matching all anomalies.  At this point, it is instructive to compare the set of conditions $(i)$ to $(iii)$ in (\ref{conditions for absence of massless spectrum}) with those that result from turning on BCF fluxes on $\mathbb T^4$, as was considered before\footnote{For the sake of completeness, we recall that the conditions (\ref{conditions for absence of massless spectrum}) are replaced  on $\mathbb T^4$ by:  
\begin{eqnarray}
N_c\ell^c-\ell^Bn_c^2\in N_c^2\mathbb Z\,,\quad N_f\ell^f-\ell^B n_f^2\in  N_f^2\mathbb Z\,,\quad \ell^B\in {\cal Q}\frac{N_c N_f}{n_cn_f}\mathbb Z\,,
\label{lattice condition}
\end{eqnarray}
where ${\cal Q}$ is the smallest integer that makes  ${\cal Q}\frac{N_c N_f}{n_cn_f}$ an integer.
 } 
  \cite{Anber:2019nze}. Although the two sets of conditions appear to be unrelated, they give the exact same no-go condition (\ref{no go conditions}).

\item However, as we shall show in the examples in Section \ref{examples}, putting the theory on a non-spin manifold can give rise to a more restrictive phase in the partition function, and hence, imposes more constraints on the IR TQFT that accompanies the composites. 

\item As in \cite{Cordova:2019uob,Anber:2019nfu}, we can also turn on a $SU(N_f)$ invariant  mass term that breaks $SU(N_f)_L\times SU(N_f)_R$ down to the diagonal  vector  subgroup. We will take the mass to be smaller than the strong-coupling scale of the theory and also introduce a $\theta$ parameter. Now, we examine how the partition function transforms under a shift of $\theta$ by multiples of $2\pi$, i.e.,~we ask whether the theory suffers a $\theta$-periodicity anomaly. To this end, we introduce, in addition to the $\theta$ term,   general  background field dependent counter terms. The topological part of the Lagrangian becomes ${\cal L}_{top.}= \theta Q^c+\Theta_f Q^f+ \Theta_B Q^B+\frac{\Theta_G}{2}Q^G$, where the coefficients of the counterterms, $\Theta_f,\Theta_B,\frac{\Theta_G}{2}$, are general real numbers. They can, however, depend on $\theta$ and we demand that they shift by $2\pi \Z$ under $2\pi r$ shifts of $\theta$, so that they do not destroy the $\theta$ periodicity in  backgrounds with integer  $Q^c, Q^f, Q^B$ and even $Q^G$. In other words, we have that under  $\theta\rightarrow \theta+2\pi r$ (where $r \in \mathbb Z$),  $\Delta {\cal L}_{top.} = 2 \pi r Q^c + 2 \pi s  Q^f+ 2 \pi t Q^B+\frac{2 \pi u}{2}Q^G$, where $s,t,u \in \Z$.

Finally, we ask whether  the transformation of the counter terms can compensate for the phase of the partition function under shifts of $\theta$ in the  BCF background fluxes on $\mathbb{CP}^2$, i.e., we demand that  under $\theta\rightarrow \theta+2\pi r$, ${\cal L}_{top.}\rightarrow {\cal L}_{top.}+\Delta {\cal L}_{top.}$, with $\Delta {\cal L}_{top.}=2\pi \mathbb Z$. Carrying out this exercise, we find that the requirement $\Delta {\cal L}_{top.}=2\pi \mathbb Z$ (the absence of a $\theta$-periodicity anomaly) is met for general BCF fluxes if and only if conditions (\ref{conditions for absence of massless spectrum}) are satisfied after replacing $\ell^{c,f,B,G}\rightarrow r,s,t,u$. Therefore, the conditions that exclude massless composites  are the exact same conditions that give rise to $\theta$-periodicity anomaly: they are given, for $N_f \ge 2$, by  the same conditions (\ref{no go conditions}) found earlier in \cite{Anber:2019nfu}. The anomaly implies that as one varies $\theta$ between $0$ and $2\pi$, the IR theory should either have domain walls or an IR TQFT that saturates the anomaly. 
\end{enumerate}
%

\subsection{Examples}

\label{examples}

In this Section, we consider two examples of vector-like theories  and check whether putting them on non-spin manifolds and turning on the most general background fluxes imposes further restrictions on various scenarios for their IR dynamics.  Many aspects of what we find have been previously recognized in \cite{Cordova:2018acb,Bi:2018xvr,Wan:2018djl,Cordova:2019bsd, Cordova:2019jqi}, especially in the framework of QCD(adj), our first example below. Nonetheless, we include it in order to show how it fits in the present more general and explicit framework.

\subsubsection{QCD(adj)}

\label{qcdadj}

As a first example, we consider QCD(adj), 
an $SU(N_c)$ Yang-Mills theory endowed with $N_f$ massless Dirac flavors in the adjoint representation. The Dirac fermion is equivalent to two undotted Weyl massless fermions $\psi, \tilde\psi$, both transforming in the adjoint representation. 
The global symmetry of this theory that we shall utilize is
\begin{eqnarray}\label{adjointsymmetry}
G^{\scriptsize\mbox{Global}} \supset \frac{SU(N_f)_L\times SU(N_f)_R\times U(1)_B\times \mathbb Z_{4N_c N_f}}{\mathbb Z_{N_f}\times \mathbb Z_2}\times \mathbb Z_{N_c}^{(1)}\,,
\end{eqnarray}
where we included the $1$-form $\mathbb Z_{N_c}^{(1)}$ center symmetry that acts on Polyakov loops. 
The massless Dirac theory above is equivalent to the theory of $2 N_f$ massless Weyl adjoints\footnote{Here $i=1,..., 2N_f$ and all $\lambda^i$ are undotted $SL(2,\mathbb{C})$ spinors.}  $\lambda^i$, which has a larger global $SU(2N_f)$ chiral symmetry, containing the $SU(N_f)_L\times SU(N_f)_R\times U(1)_B$ shown above. While studying the BCF anomaly on non-spin manifolds, however, we shall make use  of the backgrounds  (\ref{fluxes in Cartan}) for the symmetry  (\ref{adjointsymmetry}). 

This class of theories has been extensively studied in the continuum \cite{Sannino:2004qp,Unsal:2007jx,Cherman:2018mya} and on the lattice  \cite{Catterall:2007yx,Hietanen:2009az,DelDebbio:2010hx,Athenodorou:2015fda,Bergner:2016hip,Bergner:2017bky,Bergner:2017gzw,Bergner:2018fxm,Bi:2019gle}, for general theoretical interest, but also  because it  includes  theories of interest for model building beyond the Standard Model. 
The usual lore is that these theories will either flow to an IR conformal field theory or break their global symmetries, including the discrete chiral symmetry $\mathbb Z_{4N_c N_f}$. However, more exotic scenarios have recently been discussed in \cite{Anber:2018tcj,Cordova:2018acb,Bi:2018xvr,Wan:2018djl,Poppitz:2019fnp}.\footnote{We stress that while comparing the results in these references to the ones given here, one should keep in mind that $N_f$ in this paper denotes the number of Dirac, not Weyl flavors. Thus, the discussion here applies to even numbers of Weyl flavors.}

 In \cite{Anber:2018tcj}, we conjectured that the theory with $N_c=2$ and a single $N_f=1$ Dirac fermion will form a massless composite,  schematically given by $ (\lambda)^3$, a doublet under the enhanced $SU(2 N_f) =  SU(2)$ flavor symmetry, accompanied by the breaking $\mathbb Z_8\rightarrow \mathbb Z_4$, due to an $SU(2)$ invariant four-fermion condensate. This IR scenario has to be supplemented by a TQFT that matches a mixed anomaly between the $0$-form discrete chiral and $1$-form center symmetries on non-spin backgrounds  \cite{Cordova:2018acb}, further studied in  \cite{Wan:2018djl,Cordova:2019bsd, Cordova:2019jqi}.

 Another exotic scenario, applicable to all $N_c$, $N_f$, is the proposal of \cite{Poppitz:2019fnp}, where the IR phase of the theory contains $(N_c^2-1) \times 2 N_f$ massless fermions (essentially providing a gauge invariant copy of the UV fermion spectrum) which can be thought of as created by operators of the form:
\begin{eqnarray}
\label{crazy composites}
{\cal O}_1^i= \mbox{tr}\left[F_{\mu\nu}\gamma^{\mu\nu}\lambda^i\right],~~\ldots ~,
{\cal O}_{N_c^2-1}^i = \mbox{tr}\left[\underbrace{F_{\mu\alpha}...F_{\rho\nu}}_{N_c^2-1}\gamma^{\mu\nu}\lambda^i\right]\,.
\end{eqnarray}
This class of composites match all the $0$-form anomalies. In addition, there is a TQFT that matches the discrete chiral-center anomaly. Clearly this is also required by the ``no-go condition" $(ii.)$ from (\ref{no go conditions}) as gcd$(N_c, n_c=N_c) = N_c > 1$. 

 It will be instructive to check whether putting QCD(adj) on $\mathbb{CP}^2$ can impose further constraints on the above IR scenarios. To this end, we first
  examine the transformation of the partition function in the UV under the $\mathbb Z_{4N_c N_f}$ discrete chiral symmetry. The index (\ref{Dirac index}) is now given by
\begin{eqnarray}
{\cal J}_D=2 N_c N_f Q^c+(N_c^2-1)Q^f+N_f(N_c^2-1)(Q^B+Q^G)\,,
\end{eqnarray}
where $Q^{c,f,B}$ are given in (\ref{fractional top charges}) after setting $n_c= 0$ and $n_f=1$. This index is always an integer for all $m^c$ and $m^f$, as can be easily checked. Then, under a $\mathbb Z_{4N_c N_f}$ transformation the partition function acquires the phase\footnote{The $U(1)_B$ background is taken to have an extra  flux $m^B \in \Z$, $F^{B}=(\frac{1}{2}+ m^B+\frac{n^f}{N_f}m^f) K$, cf.~(\ref{fluxes in Cartan}).}
\begin{eqnarray}
\nonumber
\left({\cal Z}_{UV}|_{\mathbb Z_{4N_c N_f}}\right)_{\mathbb{CP}^2}\rightarrow {\cal Z}e^{i \frac{2\pi}{2N_cN_f}\left[N_f(m^c)^2(N_c-1)+(N_c^2-1)\left(\frac{m^f(m^f+1)}{2} +   m^B m^f + N_f \frac{m^B(m^B+1)}{2} \right)\right]}\,.\\
\label{uvqcdadj}
\end{eqnarray}
Thus, $ {\cal Z}_{UV}$ transforms by a $\Z_{2 N_cN_f}$ phase  for general values of the background BCF fluxes.

Now, we first examine the IR scenario \cite{Anber:2018tcj} for $N_c=2$ and a single Dirac fermion $N_f=1$. The IR composite Dirac fermion has unit charge under $U(1)_B$ and charge $3$ under the $\Z_8$ discrete chiral symmetry.\footnote{Recall that $U(1)_B$ is really the third component of the enhanced $SU(2)$ flavor symmetry of the two-Weyl theory and that the massless fermion is an $SU(2)$ doublet.} The Dirac index in the IR is obtained by setting $Q^c=Q^f=0$ in  (\ref{Dirac index}), which gives ${\cal J}_D=\frac{1}{2}m^B(m^B+1)$. Thus, we find $\left({\cal Z}_{IR}|_{\mathbb Z_{8}}\right)_{\mathbb{CP}^2}\rightarrow e^{ i\frac{2\pi \times 3}{8}m^B(m^B+1)}$, and hence, from (\ref{uvqcdadj}) we find the ratio 
\begin{eqnarray}\label{diracadj1}
\left(\frac{{\cal Z}_{UV}|_{\mathbb Z_8}}{{\cal Z}_{IR}|_{\mathbb Z_8}}\right)_{\mathbb{CP}^2}=e^{i\frac{2\pi}{4} (m^c)^2}\,.
\end{eqnarray}
We note that on a non-spin manifold, this is a $\Z_4$ phase, while it is a $\Z_2$ phase on a  spin manifold. On $\mathbb T^4$, the computation  follows the same steps, taking  $SU(N_c)$ 't Hooft fluxes (see footnote \ref{t4thooft}), with $Q^c = {m m'\over 2}$, $Q^f = 0$, and taking $Q^B = m_b$ 
($m,m', m_b \in \Z$), we have 
 \begin{eqnarray}\label{diracadj2}
\left(\frac{{\cal Z}_{UV}|_{\mathbb Z_8}}{{\cal Z}_{IR}|_{\mathbb Z_8}}\right)_{\mathbb{T}^4}={e^{i\frac{2\pi}{4}\left(4 {mm'\over 2} + 3 m_b\right)} \over e^{i\frac{2\pi}{4} 3  m_b}} = e^{i {2 \pi \over 2 } mm'}
\end{eqnarray}

The fact that the UV and IR partition functions with massless composite fermions transform differently under $\Z_8$ means that the massless composites cannot be all there is in the IR. In particular, as (\ref{diracadj1}, \ref{diracadj2}) show, there is a mixed anomaly between the discrete chiral and center symmetries (the 't Hooft fluxes $m, m', m^c$) which cannot be matched by the IR fermions.
This was already recognized in \cite{Anber:2018tcj}, where it was proposed that there is spontaneous breaking of the chiral symmetry, $\Z_8 \rightarrow \Z_4$, by a four-fermion condensate $\langle{\rm det} \lambda^i \lambda^j \rangle$\footnote{The determinant is taken in the 2-dimensional space of Weyl flavors.} and   that domain walls, via a TQFT  coupled to the background fields and describing the two $\Z_8 \rightarrow \Z_4$ vacua, match the mixed discrete-chiral center anomaly.

Consider, however, a chiral   transformation in the unbroken $\Z_4$. A look at (\ref{diracadj1}) and (\ref{diracadj2}) shows that an unbroken-$\Z_4$ transformation  (a $\Z_8$ transformation applied twice) generates no phase on $\mathbb T^4$, but does give rise to a $\Z_2$ phase on $\mathbb{CP}^2$. The DW theory, however, is blind\footnote{A theory with two vacua and  domain walls between should be described, in the IR, by a $\Z_2$ TQFT with  Euclidean Lagrangian $i {2 \over 2 \pi}\int \phi^{(0)} (d a^{(3)} + \ldots)$, see \cite{Hidaka:2019mfm} for a recent discussion. Here, $\phi^{(0)}$ and $a^{(3)}$  are compact 
0-form and 3-form gauge fields ($d \phi^{(0)}$ and $d a^{(3)}$ have periods  $2 \pi \Z$ when integrated over appropriate cycles) and the dots denote  background field couplings. Under the action of the broken $\Z_8$ generators, $\phi^{(0)}$ shifts by $\pi$, but is inert under the unbroken $\Z_4$ generators.}  to the unbroken $\Z_4$ group and only matches the anomalies for the broken symmetries, generated by odd powers of $e^{i {2 \pi \over 8}}$.
 Thus to match the anomaly of the unbroken $\Z_4$ group \cite{Cordova:2018acb}, the scenario   proposed in \cite{Anber:2018tcj}   has to be modified. The need for such modification is only visible---as (\ref{diracadj1}, \ref{diracadj2}) show---when the theory is placed in consistent non-spin backgrounds.
  It was argued that one would need to supplement the IR  with an extra emergent TQFT  and an explicit construction of this TQFT as an emergent $\Z_2$ gauge theory matching the anomaly of the unbroken $\Z_4$ on  non-spin manifolds (giving rise to the $\Z_2$ phase) was given  \cite{Cordova:2018acb,Wan:2018djl,Cordova:2019bsd, Cordova:2019jqi}.

Next, we examine the scenario of \cite{Poppitz:2019fnp}. The massless composites (\ref{crazy composites}) have unit $U(1)_B$ and $\Z_{4 N_c N_f}$ charges, hence the index in the IR is  ${\cal J}_D=(N_c^2-1)\left[ Q^f+N_F(Q^B+Q^G)\right] = (N_c^2-1) {m_f (m_f+1) \over 2}$. Thus, we
 find, proceeding as above and taking $m^B=0$ with no loss of generality, that 
 \begin{eqnarray}\label{cp2ryttov}
  \left(\frac{{\cal Z}_{UV}|_{\mathbb Z_{4N_fN_c}}}{{\cal Z}_{IR}|_{\mathbb Z_{4N_fN_c}}}\right)_{\mathbb{CP}^2}=e^{ i\frac{2\pi \left(m^c\right)^2}{2}\left(1-\frac{1}{N_c}\right)}\,.
 \end{eqnarray}
 Again, we find that this phase is half the phase one obtains from the mixed anomaly between the discrete chiral and center symmetries on spin manifolds. Ref.~\cite{Poppitz:2019fnp} proposed that a higher-dimensional condensate breaks $\mathbb Z_{4N_fN_c}  \rightarrow \mathbb Z_{4N_f} $, but as in the above $N_f=1, N_c=2$ example, this is not sufficient to match the anomaly of the unbroken $\Z_{4 N_f}$ symmetry on $\mathbb{CP}^2$ (it is clear, by applying (\ref{cp2ryttov}) $N_c$ times, that this is a $\Z_2$-valued anomaly).  Thus, we conclude that an additional emergent TQFT, argued  to also be an emergent $\Z_2$ gauge theory  \cite{Cordova:2019bsd}, has to exist in the IR to match the anomaly of the unbroken $\Z_{4 N_f}$ symmetry on $\mathbb{CP}^2$. 
 
 To summarize, in both  scenarios \cite{Anber:2018tcj,Poppitz:2019fnp}, the IR theory consists of three decoupled sectors: massless composite fermions, a $\Z_{N_c}$ TQFT due to the spontaneous chiral symmetry breaking (with $N_c$ vacua and domain walls), and an  emergent topological $\Z_2$ gauge theory. Here, we shall not speculate on the likelihood of this scenario and simply refer the reader to \cite{Bi:2019gle} for the up-to-date status of the lattice studies.

\subsubsection{$\mathbf{SU(6)}$ with a Dirac fermion  in the two-index anti-symmetric representation}

\label{su6}

As our  second study of the new anomaly, we consider $SU(N_c=6)$ vector-like theory with a single Dirac spinor with ${\cal R}$ taken to be the two-index antisymmetric representation ($N$-ality $n_c = 2$). We denote its two undotted Weyl-fermion components as $\psi, \tilde\psi$, transforming in ${\cal R}$ and $\overline{\cal R}$, respectively.
Recalling (\ref{globalG}), the global symmetry of this theory is 
\begin{eqnarray}
G^{\mbox{global}}=\frac{U_B(1)\times \mathbb Z_{8}}{\mathbb Z_3\times\mathbb Z_2}\times \mathbb Z^{(1)}_2\,,
\end{eqnarray}
where we modded by the $\mathbb Z_3$, the discrete group that acts faithfully on fermions,  and the $\mathbb Z_2$ subgroup of the Lorentz group, while the $1$-form center symmetry $\mathbb Z^{(1)}_2$ should be understood as acting on topologically nontrivial Wilson loops.

A possible phase of the theory  is one where a bilinear fermion condensate $\langle \tilde\psi \psi\rangle$ forms. This condensate preserves the vectorlike $U(1)_B$ but breaks $ \mathbb Z_{8}$ down to $\mathbb Z_2$. The theory is gapped and in the deep IR  the anomaly is matched by a $\Z_4$ TQFT describing the four ground states of the theory. This number of vacua is consistent with the constraints on gapped phases of such theories recently derived in \cite{Cordova:2019jqi}. This is also the breaking pattern expected when the theory is coupled to an axion \cite{Anber:2020xfk}.

 In what follows, we study the viability of a more exotic scenario for the IR physics, namely the possibility  to match the anomalies via a single massless composite Dirac fermion of the form\footnote{Derivative and field strength insertions may be required in the precise definition of ${\cal O}$, $\tilde{\cal O}$. These, however, do not affect the $U(1)_B$ and $\Z_8$ quantum numbers of relevance here.} ${\cal O}\sim\left(\psi\right)^3$, $\tilde{\cal O}\sim (\tilde\psi)^3$, which has charge $3$ under both $U(1)_B$ and $\Z_8$. It is a simple exercise to check that all the $0$-form anomalies are matched by the ${\cal O}$ composite. Using (i) in (\ref{conditions for absence of massless spectrum}), ignoring (ii), (iii), and (iv) since we are dealing with a single Dirac fermion,   one can easily show that there is no integer $\ell^c$ that satisfies (\ref{conditions for absence of massless spectrum}). Hence, additional IR data to a massless fermions spectrum is needed.

Next, we check whether ${\cal O}$ matches the BCF 't Hooft anomaly on $\mathbb{CP}^2$. We will also compare the result with that of the BCF anomaly on a spin manifold. To this end, let us examine the change of the partition function under a global $\mathbb Z_8$ chiral transformation in the background of the BC fluxes on $\mathbb{CP}^2$. From (\ref{UV Z}), using (\ref{fractional top charges}, \ref{grav charge})  with $m_f=0$, and recalling that the anomaly is twice the Dirac index (\ref{Dirac index}), ${\cal J}_D=\frac{5}{2}m^c(m^c+1)$, we find in the UV:
\begin{eqnarray}
\left({\cal Z}_{UV}|_{\mathbb Z_8}\right)_{\mathbb{CP}^2}\rightarrow  {\cal Z}e^{i {2 \pi \over 8} 2 {\cal J}_D} ={\cal Z}e^{i 2\pi  \frac{m^c(m^c+1)}{8}}\,.
\end{eqnarray}
 In the IR, the Dirac index\footnote{The IR composite only couples to the gravitational and baryon number backgrounds (\ref{fluxes in Cartan}), hence $Q^c = Q^f = 0$. In addition, since the baryon charge of $\cal{O}$ is $3$,  the formula for the index (\ref{Dirac index}) has to be modified by multiplying $Q_B$ by $3^2$ and taking $\mbox{dim}_{{\cal R}^f}\mbox{dim}_{{\cal R}^c}=1$.}   for the composite ${\cal O}$ is ${\cal J}_D=1+\frac{m^c}{2}\left(m^c+3\right)$, thus we  find
\begin{eqnarray}
\left({\cal Z}_{IR}|_{\mathbb Z_8}\right)_{\mathbb{CP}^2}\rightarrow  {\cal Z}e^{i {2 \pi \over 8} 3 \times 2 {\cal J}_D} = {\cal Z}e^{i2\pi\frac{3}{4}\left(1+\frac{m^c(m^c+3)}{2}\right)}\,.
\end{eqnarray}
Therefore, the ratio between the $\Z_8$ chiral transformations of the partition function in the UV and IR theories in the same BC background (\ref{fluxes in Cartan}) is
\begin{equation}\label{mismatchsu6}
\left(\frac{{\cal Z}_{UV}|_{\mathbb Z_8}}{{\cal Z}_{IR}|_{\mathbb Z_8}}\right)_{\mathbb{CP}^2}=e^{i\frac{2\pi}{4}(-1+(m^c)^2)}~.
\end{equation}
If the massless composite $\cal{O}$ matches all anomalies, the phase on the r.h.s. of (\ref{mismatchsu6}) should be unity for all values of the $SU(N_c)$ 't Hooft fluxes $m^c$. Clearly, this is not the case and (\ref{mismatchsu6}) implies 
  that there is a  $\frac{\pi}{2}$ phase mismatch between the UV and IR 't Hooft anomalies on $\mathbb{CP}^2$. This phase is obtained even if we completely turn off the  $SU(N_c)$ 't Hooft fluxes by setting $m^c=0$, hence the anomaly is solely due to putting the theory on a non-spin manifold, i.e., there is a mixed anomaly between the $0$-form $\mathbb Z_8$ discrete chiral symmetry and the $U(1)_B-\mbox{gravity}$ background required to put the theory on $\mathbb{CP}^2$. The mismatch (\ref{mismatchsu6}) indicates  that a single composite in the IR cannot by itself match  this mixed anomaly. In addition to the composite, the theory has  to be supplemented by  partial breaking of $\mathbb Z_8$ and/or an  IR TQFT.  

It is also important to compare the situation with the BCF anomaly on a spin manifold. One can repeat the above exercise on $\mathbb T^4$ to find,  in the background of BC fluxes (recall (\ref{diracadj2})) \begin{equation}\label{t4anomalysu6}
 \left(\frac{{\cal Z}_{UV}|_{\mathbb Z_8}}{{\cal Z}_{IR}|_{\mathbb Z_8}}\right)_{\mathbb T^4}=e^{i\pi m m'}~,
 \end{equation} instead of (\ref{mismatchsu6}) on $\mathbb{CP}^2$. The $\pi$ phase mismatch can also be obtained as the result of a mixed anomaly between $\mathbb Z_8$ and the $1$-form $\mathbb Z_2^{(1)}$ center  symmetry \cite{Bolognesi:2019fej}. In both $\mathbb{CP}^2$ and $\mathbb T^4$ cases (\ref{mismatchsu6}, \ref{t4anomalysu6}) we find that one needs to supplement the theory with an emergent IR TQFT in order to match the   phases in (\ref{mismatchsu6}, \ref{t4anomalysu6}). 
The $\mathbb T^4$ UV/IR phase mismatch (\ref{t4anomalysu6}), for the broken $\Z_8$ generators, could be due to domain walls from the spontaneous breaking $\Z_8 \rightarrow \Z_4$ by a $(\tilde\psi \psi)^2$-condensate (recall also that $\Z_2 \in \Z_4$ is fermion number). This, however would not match the nontrivial  $\Z_2$-valued anomaly in the unbroken-$\Z_4$ transformation of the partition function  on a non-spin manifold (\ref{mismatchsu6}). Thus, we conclude that, once again, putting the theory on a non-spin manifold
gives more constraints on the IR physics,  by requiring  an extra TQFT to match the anomaly of the unbroken $\Z_4$ symmetry on $\mathbb{CP}^2$ (the phase to be matched is, again, a $\Z_2$ phase). The results of  \cite{Cordova:2019bsd} imply that such a $\Z_4$ and $\Z_2^{(1)}$-center symmetric TQFT exists: the anomaly inflow action is nontrivial if one assumes $\Z_8$ and $\Z_2^{(1)}$ unbroken symmetries (precluding the existence of a symmetric gapped phase  \cite{Cordova:2019bsd}), but trivializes for the case of unbroken $\Z_4$ and $\Z_2^{(1)}$. See also the discussion of the more general case near eq.~(\ref{inflow action}) in the following Section.

\subsubsection{$\mathbf{SU(4k+2)}$ with fermions in the two-index (anti)-symmetric representation}

\label{su4k2}

Here, we generalize the $SU(6)$-theory analysis to $SU(4k+2)$ with a single Dirac fermion in the two-index symmetric (S) or anti-symmetric (AS) representation\footnote{Notice that $SU(4k)$ with fermions in the two-index S or AS does not admit color-singlet fermions in the IR. Hence, we exclude this case from our discussion.}.
The conclusion, with regards to an IR phase with composite massless fermions, is essentially the same as in the $SU(6)$ theory of Section \ref{su6}. Below, we give the details for completeness.

 We turn on color and baryon-number fluxes and use (\ref{Dirac index}) to calculate the Dirac index in the UV. Recalling that $Q^c=\frac{(m^c)^2}{2}\left(1-\frac{1}{4k+2}\right)$, $Q^B=\frac{1}{2}\left(\frac{1}{2}+\frac{2m^c}{4k+2}\right)^2$, $T_{\scriptsize\mbox{S}, \mbox{AS}}=4k+2\pm 2$, and $\mbox{dim}_{\scriptsize\mbox{S}, \mbox{AS}}=\frac{1}{2}(4k+2)(4k+2\pm1))$,  we find
\begin{eqnarray}
{\cal J}_D^{UV}=\left\{\begin{array}{lr}\frac{m^c}{2}(3+5m^c+4k(1+m^c))& \mbox{S}\,, \\ \frac{m^c}{2}(m^c+1)(4k+1)& \mbox{AS}\,, \end{array}  \right.
\end{eqnarray}
from which one can readily find that the partition function receives the following phases upon performing a discrete chiral symmetry transformation $\mathbb Z_{2(4k+2\pm2)}$:
\begin{eqnarray}
\left({\cal Z}_{UV}|_{\mathbb Z_{2(4k+2\pm2)}}\right)_{\mathbb{CP}^2}\rightarrow {\cal Z}e^{i\frac{2\pi}{4k+2\pm2}{\cal J}_D^{UV}}\,.
\end{eqnarray}

As above, we focus on   the anomaly constraints on an exotic scenario for the IR physics.
We assume that the $0$-form anomalies are saturated in the IR by a set of massless composites. This can be achieved in the AS case by a single  composite  ${\cal O}\sim(\psi)^{2k+1}$ and single anti-composite $\tilde {\cal O}\sim(\tilde\psi)^{2k+1}$, while in the S case we need\footnote{To match the $0$-form anomalies involving $\Z_{2(4k+4)}$.} $3+4k$ composites ${\cal O}\sim(\psi)^{2k+1}$ and anti-composites $\tilde {\cal O}\sim(\tilde\psi)^{2k+1}$, possibly with appropriate insertions of derivatives and/or gluonic fields. Since all the IR composites are color singlets, only the baryon flux will contribute to the Dirac index: 
\begin{eqnarray}
{\cal J}_D^{IR}=(2k+1)^2Q^B-\frac{1}{8}=\frac{1}{2}\left[k(k+1)+m^c(1+2k+m^c)\right]
\end{eqnarray}
for each of the symmetric and anti-symmetric Dirac composites, and we used the fact that the $U(1)_B$ charges of the composites is $2k+1$. Using this information, we obtain the following phases in the partition function upon performing a discrete chiral transformation:
\begin{eqnarray} 
\left({\cal Z}_{IR}|_{\mathbb Z_{2(4k+2\pm2)}}\right)_{\mathbb{CP}^2}\rightarrow {\cal Z}\times \left\{ \begin{array}{lr} e^{i 2\pi\frac{(2k+1)(3+4k){\cal J}_D^{IR}}{4k+4} } &\mbox{S} \\ e^{i 2\pi\frac{(2k+1){\cal J}_D^{IR}}{4k} } & \mbox{AS} \end{array} \right.\,,
\end{eqnarray}
where we used the fact that we need $3+4k$ composites in the symmetric case. Finally, after some algebra  we obtain the ratios:
\begin{eqnarray}
\left(\frac{{\cal Z}_{UV}|_{\mathbb  Z_{2(4k+2\pm 2)}}}{{\cal Z}_{IR}|_{\mathbb  Z_{2(4k+2\pm2)}}}\right)_{\mathbb{CP}^2}= \left\{ \begin{array}{lr}e^{ i2\pi \frac{-3k-10k^2 - 12k m^c + 2 (m^c)^2}{8}} & \mbox{S}\,\\ e^{i2\pi \frac{-1 - 2 k^2 + 2 (m^c)^2 - k (3 + 4 m^c)}{8}} &\mbox{AS}\end{array} \right..
\label{general phase}
\end{eqnarray}
This phase mismatch between the UV and IR implies that turning on BC fluxes on $\mathbb {CP}^2$ rules out the set of composites as the sole spectrum in the IR.  For  the S  case, we  obtain a $\Z_8$-valued anomaly on $\mathbb{CP}^2$  for odd values of $k$, and a  $\Z_4$-valued one for even values of $k$, while for the AS case we obtain a $\Z_4$ phase for odd-$k$ and a $\Z_8$ phase for even-$k$.

Before we continue with studying the implications of (\ref{general phase}), let us contrast the situation on $\mathbb {CP}^2$ with that on $\mathbb T^4$. In the latter case we can turn on general color and baryon fluxes in the $1$-$2$ and $3$-$4$ planes: $Q^c=m_{12}^cm_{34}^c\left(1-\frac{1}{4k+2}\right), Q^B=\frac{4m_{12}^cm_{34}^c}{(4k+2)^2}$. Then, the Dirac index in the UV is given by
\begin{eqnarray}
{\cal J}_{D}^{UV}=m_{12}^cm_{34}^c(4k+3\pm 2)\,,
\end{eqnarray}
for the S and AS cases, respectively. In the IR the composites are color singlets, they have   charge $2k+1$ under $U(1)_B$, and therefore, the index is
\begin{eqnarray}
{\cal J}_{D}^{IR}=(2k+1)^2\frac{4m_{12}^cm_{34}^c}{(4k+2)^2}=m_{12}^cm_{34}^c.
\end{eqnarray}
Repeating the above steps, we obtain the following phases upon performing a $\mathbb Z_{2(4k+2\pm 2)}$ discrete chiral transformations in the BC fluxes:
\begin{eqnarray} \label{t4anomaly}
\left(\frac{{\cal Z}_{UV}|_{\mathbb  Z_{2(4k+2\pm 2)}}}{{\cal Z}_{IR}|_{\mathbb  Z_{2(4k+2\pm2)}}}\right)_{\mathbb T^4}=\left\{ \begin{array}{lr}e^{i \pi {m_{12}^cm^c_{34}}}\, & \mbox{S}\,\\ e^{i\pi{m_{12}^cm^c_{34}}} & \mbox{AS}\end{array} \right.\,.
\end{eqnarray}
Here, the phase we obtain is the exact same $\Z_2$ phase one encounters from the discrete-chiral/$1$-form $\mathbb Z_2$-center anomaly.

The symmetry breaking scenario consistent with the above massless composite spectrum is as follows. For the case of symmetric tensor (S) representation, we assume a nonvanishing $(\psi \tilde\psi)^{2k+2}$ condensate (with all other condensates  zero) breaking the chiral symmetry $\Z_{2(4k+4)} \rightarrow \Z_{4k+4}$. The anomaly inflow 5d action has the form
\begin{equation}\label{inflow action}
e^{i {2 \pi \over 2} \int_{M_5} {2 (4k+4) A^{(1)} \over 2 \pi} \wedge {2 B^{(2)}\over 2 \pi} \wedge {2 B^{(2)}\over 2 \pi} } ~,
\end{equation}
with $A^{(1)}$ a 1-form gauge field for $\Z_{2(4k+4)}$ and $B^{(2)}$ a 2-form gauge field for the $\Z_2^{(1)}$ center symmetry.\footnote{The normalization and transformation properties of $A^{(1)}$ and $B^{(2)}$ are as in \cite{Gaiotto:2014kfa,Gaiotto:2017yup,Gaiotto:2017tne}.} The chiral variation of (\ref{inflow action}) reproduces the $\Z_2$-valued mixed anomaly (\ref{t4anomaly}). In addition,  (\ref{inflow action}) evaluates to $e^{i \pi}$ on $\S^1 \times \S^2 \times \S^2$, thus, according to \cite{Cordova:2019bsd} no $\Z_{2(4k+4)}$- and $\Z_2^{(1)}$-symmetric unitary TQFT exists to match this anomaly, implying that the symmetry has to suffer at least partial breakdown. However, when 
$\oint {2 (4k+4) A^{(1)} \over 2 \pi} = 2$, i.e. with the background  restricted to the unbroken $\Z_{4k+4}$, the expression (\ref{inflow action}) evaluates to unity and a symmetric TQFT matching the unbroken symmetries anomaly is not excluded.\footnote{The recent ref.~\cite{Thorngren:2020aph} asserts that all symmetric TQFTs not excluded by \cite{Cordova:2019bsd} do in fact exist.}
 A similar scenario with  $\Z_{8k} \rightarrow \Z_{4k}$ symmetry breaking, due to a nonzero $(\psi \tilde\psi)^{2k}$ condensate, holds for the AS case. 
 
 As in the composite-fermion QCD(adj) scenarios discussed in the previous Section, there are three decoupled sectors in the IR: massless composite fermions, domain walls and multiple vacua due to the symmetry breaking, and a TQFT to match the anomaly of the unbroken chiral symmetry. As before, we shall not dwell on the likelihood of these exotic IR phases appearing in the nonabelian gauge theories under consideration. 
 
\subsection{Comments on future studies}

In this Section, we studied a few examples illustrating the utility of the mixed chiral/BCF anomaly on non-spin backgrounds. Our main focus was on exotic phases where massless composite fermions saturate  the ``traditional" 0-form 't Hooft anomalies. The main   lesson we take is that the new generalized 't Hooft anomalies on both spin and non-spin manifolds yield further constraints. 

 It is clear that generalized 't Hooft anomalies will also have implications on the physics of   ``vanilla" phases where fermion bilinears obtain expectation values maximally breaking the chiral symmetries. As the analysis   \cite{Cordova:2018acb}  of $SU(2)$ QCD(adj) with a single Dirac flavor showed, the structure of the IR theory, its domain walls, and confining strings can reflect the anomalies in an intricate way. It would be interesting to understand the implications of  anomaly matching for similar phases in more general theories, including chiral theories or the ones studied in \cite{Anber:2019nfu}.
Constructing the IR TQFTs that must accompany the various exotic phases mentioned here is also of interest (we also note that their UV origin remains mysterious). Anomalies should also have implications for the finite temperature phase structure, as in
 \cite{Gaiotto:2017yup,Komargodski:2017smk,Shimizu:2017asf,Anber:2018jdf,Anber:2018xek}.

\bigskip

{\bf {\flushleft{Acknowledgments:}}}  We thank Georg Bergner and Andreas Wipf  for many interesting discussions. MA  gratefully acknowledges the hospitality at the University of Jena and the University of Toronto.   MA is supported by the NSF grant PHY-1720135. EP is supported by a Discovery Grant from NSERC.

\bigskip

\appendix 

\section{Some useful formulae for $\mathbb{CP}^2$}
\label{CP2 space}

In this Appendix, we review important facts about the complex projective space $\mathbb{CP}^2$. Our notation largely follows  \cite{Gibbons:1978zy,Eguchi:1980jx}.
$\mathbb{CP}^2$ is the set of  lines in the three-dimensional complex space, $\mathbb C^3$, passing through the origin. $\mathbb{CP}^2$ can be described by the complex coordinates $\Xi=(\xi_1,\xi_2, \xi_3)\neq (0,0,0)$ (here $\xi_{1,2,3}\in \mathbb C$) modulo the identification $\Xi\equiv \lambda \Xi$ for any complex number $\lambda \neq 0$. One can cover $\mathbb{CP}^2$ with three patches $U_i$  ($i=1,2,3$, where $U_i$ covers $\xi_i \ne 0$)  such that the transition functions on the overlap $U_i\cap U_j$ are holomorphic.
$\mathbb{CP}^2$ is a K\"{a}hler manifold, with a K\"{a}hler $2$-form given by
\begin{eqnarray} \label{kahlertwoform}
K= i \; \partial\wedge\bar\partial \;{\cal K}\,,
\end{eqnarray}
where $\partial$ is defined as $\partial f\equiv \sum_{\alpha}\frac{\partial f}{\partial z^\alpha}dz^\alpha$ (and similarly for $\bar \partial$) and ${\cal K}$ is the K\"{a}hler potential:
\begin{eqnarray}\label{kahler1}
{\cal K}=\log\left(1+\sum_{\alpha=1}^2z^\alpha\bar z^\alpha\right)\,,
\end{eqnarray}
where $z^{1,2}$ cover one of the patches $U_i$. Taking $z^1 \equiv \xi^1/\xi^3, z^2 \equiv \xi^2/\xi^3$, this is the $U_3$ patch with $\xi_3 \ne 0$. At the points $\xi^3=0$ in $\mathbb{CP}^2$, we have $(\xi^1, \xi^2) \equiv \lambda (\xi^1, \xi^2)$, i.e. a two-sphere  $\S^2=\mathbb{CP}$$^1$.  In the coordinates used in (\ref{kahler1}), the $\S^2$ is  at  $|z|^\alpha \rightarrow \infty$. (This is also clear from  the explicit expression for the metric (\ref{study Fubini metric}), shown in polar coordinates in (\ref{metricpolar}).) 

The K\"{a}hler $2$-form (\ref{kahlertwoform}) is closed, $dK=0$, and co-closed, $\delta K=0$, and is  associated to the metric tensor $g_{\alpha\bar\beta}$:
\begin{eqnarray}
K= i \; g_{\alpha\bar \beta}dz^\alpha\wedge d{\bar z}^\beta\,.
\end{eqnarray}
Therefore, we immediately find
\begin{eqnarray}
g_{\alpha\bar\beta}=\frac{\delta_{\alpha\beta}}{1+\sum_{\alpha=1}^2z^\alpha\bar z^\alpha}-\frac{\bar z^\alpha z^\beta}{\left(1+\sum_{\alpha=1}^2z^\alpha\bar z^\alpha\right)^2}\,.
\end{eqnarray}
Now, one can set $z^1=x+iy$ and $z^2=z+it$ to find that the metric on $\mathbb{CP}^2$ can be written in the Fubini-Study form:
\begin{eqnarray}
ds^2=g_{\alpha\bar\beta}dz^\alpha d\bar z^\beta= \frac{dr^2+r^2\sigma_z^2}{(1+r^2)^2}+\frac{r^2\left(\sigma_x^2+\sigma_y^2\right)}{1+r^2}\,,
\label{study Fubini metric}
\end{eqnarray}
where $r^2=x^2+y^2+z^2+t^2$ and $\sigma_{x,y,z}$ are the left-invariant $1$-forms on the manifold of the group $SU(2)= \mathbb S^3$, obeying  $d \sigma_x = 2  \sigma_y \wedge \sigma_z$ (plus cyclic). The latter are given in terms of the $x,y,z,t$ coordinates by:
\begin{eqnarray}
 \sigma_x = \frac{-tdx-zdy+ydz+xdt}{r^2},~ \sigma_y=\frac{zdx-tdy-xdz+ydt}{r^2}~, \sigma_z = \frac{-ydx+xdy-tdz+zdt}{r^2}. \nonumber
\end{eqnarray}

For our explicit calculations of Appendix \ref{gauge fields and fermions}, we introduce   polar coordinates $r,\theta,\phi,\psi$ 
\begin{eqnarray}
  \label{polar1} 
  z_1 = x+ i y = r\cos\frac{\theta}{2} \; e^{i {\psi + \phi \over 2}} ~,~~ 
  z_2 = z + i t  = r\sin\frac{\theta}{2} \; e^{i {\psi - \phi \over 2}}~,
\end{eqnarray}
where $0\leq r<\infty$, $0\leq \theta<\pi$, $0\leq \phi<2\pi$, $0\leq \psi<4\pi$. The $1$-forms $\sigma_{x,y,z}$ are now
\begin{eqnarray}
\label{polar2}
\sigma_x=\frac{-\cos\psi\sin\theta d\phi+\sin\psi d\theta}{2}, ~
\sigma_y=-\frac{\cos\psi d\theta +\sin\theta \sin\psi d\phi}{2} ,~ 
\sigma_z=\frac{d\psi+\cos\theta d\phi}{2}.~~ \end{eqnarray} 
One also can write the metric in terms of the vierbein $1$-forms as
$
ds^2=e^ae^b\eta_{ab}\,,
$
where $\eta_{ab}$ is the flat Euclidean metric. Then, by inspecting (\ref{study Fubini metric}) one immediately finds:
\begin{eqnarray} \label{vierbein}
e^0=\frac{dr}{1+r^2}\,,\quad e^1=\frac{r\sigma_x}{\sqrt{1+r^2}}\,,\quad e^2=\frac{r\sigma_y}{\sqrt{1+r^2}}\,,\quad e^3=\frac{r\sigma_z}{1+r^2}\,.
\end{eqnarray}
In terms of the vierbein (\ref{vierbein}), the K\"{a}hler $2$-form (\ref{kahlertwoform}) is 
\begin{eqnarray}
K=2\left(e^0\wedge e^3+e^1\wedge e^2\right)=\frac{r}{(1+r^2)^2}dr\wedge \left(d\psi+\cos\theta d\phi\right)-\frac{1}{2}\frac{r^2}{1+r^2}\sin\theta d\theta\wedge d\phi,~~~~
\label{2-form in polar}
\end{eqnarray}
from which one can see that $K$ is anti-self-dual $\star K=-K$ ($\epsilon_{1230}=1$). We use the K\" ahler form  $K$ in polar coordinates in the calculations of fluxes and topological charges in Appendix \ref{gauge fields and fermions}.  In particular, note that 
$\int_{\mathbb{CP}^2} K\wedge K=\frac{8 \pi^2}{2}$.

The Fubini-Study metric (\ref{study Fubini metric}), explicitly written using polar coordinates (\ref{polar1}), is
\begin{equation}\label{metricpolar}
ds^2 = {d r^2 \over (1+r^2)^2} + {r^2\over 4(1+r^2)^2} (d \psi + \cos\theta d\phi)^2 + {r^2 \over 4 (1 + r^2)} (d \theta^2 + \sin^2\theta d \phi^2)~.
\end{equation}
To study the points at $r \rightarrow \infty$, one can introduce a new coordinate $u=1/r$ and observe that at $u=0$ there is a $\S^2$ of area $\pi$ (the metric is well behaved at $u=0$; the singularity apparent in the first two terms of (\ref{metricpolar}) at $1/r = u \rightarrow 0$ is only a coordinate one, see \cite{Gibbons:1978zy}).

 The Ricci tensor of the Fubini-Study metric (\ref{metricpolar}) is $R_{ab}=6 \delta_{ab}$, so it is a solution of the Einstein's equation $R_{ab}-\frac{1}{2}\delta_{ab}R=-\Lambda \delta_{ab}$ with the energy-momentum tensor being that of a cosmological constant $\Lambda=+6$. This holds for  the form of $\cal{K}$ given in (\ref{kahler1}), with dimensionless coordinates  $z^\alpha$. If, instead of (\ref{kahler1}),  we take ${\cal K}={6 \over \Lambda}\log\left(1+ {\Lambda \over 6}\sum_{\alpha=1}^2z^\alpha\bar z^\alpha\right)$, we shall find $R_{ab} =   \Lambda \delta_{ab}$, for arbitrary $\Lambda$. 
 
 Thus the compact manifold $\mathbb{CP}^2$ has a size scaling as $\Lambda^{-{1\over 2}}$. It can be taken to have any size, in particular it can be larger than $\Lambda_{QCD}^{-1}$, the inverse strong-coupling scale of the gauge theory.  Taking $\Lambda \rightarrow 0$ approaches an infinite volume limit. As in the $\mathbb T^4$ case, this is the limit of interest from the point of view of constraining infinite volume nonperturbative dynamics via anomaly matching.

\section{Gauge fields and fermions on $\mathbb{CP}^2$}
\label{gauge fields and fermions}

In order to turn on a $U(1)$ gauge field (which can be embedded into $SU(N_c)$, see below)  of two-form strength $F$ on $\mathbb{CP}^2$, one needs to ensure that the field will not backreact on the manifold, and hence, destroy $\mathbb{CP}^2$. This can be achieved by demanding that $F$  is an (anti)self-dual $2$-form field, since in this case the field has a vanishing energy-momentum tensor.\footnote{The kinetic term is $\int_{\mathbb{CP}^2}F\wedge\star F$, which, using (anti) self-duality of $F$, becomes $\pm \int_{\mathbb{CP}^2}F\wedge F$.  The latter is a metric-independent topological term, and hence, its energy-momentum tensor vanishes identically.}  Therefore,  the simplest way to find a consistent solution of the Einstein-Maxwell  equations on $\mathbb{CP}^2$ is by writing $F$ in terms of the K\"{a}hler $2$-form as $F=CK$ for some constant $C\in \mathbb R$. Below, we will see that defining spinors on $\mathbb{CP}^2$ demands that $C$ be quantized in half-integer units.

  It is well known that fermions are ill-defined on $\mathbb{CP}^2$; we say that $\mathbb{CP}^2$ is a non-spin manifold. Briefly,\footnote{For more detail see \cite{Geroch:1968zm,Geroch:1970uv,Hawking:1977ab}.} to  see that spinor fields $\Psi$ are not globally well defined,  one considers a family of closed contours $\gamma(s)$, with $s \in [0,1]$ parameterizing the different contours. This family of contours wraps the $\S^2$ in $\mathbb{CP}^2$, such that $\gamma(0)$ and $\gamma(1)$ are the trivial contours. Then one 
considers the parallel transport of tetrads, and the corresponding uplift to spinors, along each contour belonging to this family. The $SO(4)$ holonomies corresponding to parallel transporting tetrads along the family $\gamma(s)$, considered as a function of $s$, form a closed non-contractible loop in $SO(4)$ (recall that $\gamma(0)$ and $\gamma(1)$ are trivial contours). Correspondingly, the uplift of the $SO(4)$ holonomies (for the $s=0$ and $s=1$ curves) to its double cover $Spin(4)$, responsible to transporting the spinors, differ by minus sign. Schematically, one obtains 
\begin{equation}
\label{sphere}
\Psi(s=1)=e^{i\pi}\Psi(s=0) ,
\end{equation} showing  the global inconsistency (recalling that $\gamma(0)$ and $\gamma(1)$ are both the trivial contour) in defining spinors.\footnote{In a more mathematical language, the second  Stiefel-Whitney class of    $\mathbb{CP}^2$ is non-zero, indicating that there is a sign ambiguity when spinors are parallel-transported around some paths in $\mathbb{CP}^2$ \cite{Eguchi:1980jx}.} 

One can also see 
the  problem of formulating spinors on $\mathbb{CP}^2$ by computing the index of a Dirac spinor on $\mathbb{CP}^2$:
\begin{eqnarray}
{\cal J}_D= \frac{1}{192\pi^2}\int_{\mathbb{CP}^2}\mbox{tr}\left[R\wedge R\right]=-\frac{1}{8}\,.
\label{gravity index}
\end{eqnarray}
The  fractional value  $1/8$ one obtains for an integer-valued quantity (the Dirac index) is another manifestation of the failure of $\mathbb{CP}^2$ to accommodate spinor fields.

One can define spinor fields on $\mathbb{CP}^2$ if one turns on a $U(1)$ gauge bundle that eats up the $i\pi$ phase in (\ref{sphere}), which renders the spinors well-defined \cite{Hawking:1977ab}. In this case one finds that the $e^{i \pi}$ factor in (\ref{sphere}) gets modified to:
\begin{eqnarray} \label{quantization1}
e^{i2\pi\left(\frac{1}{2}+e\oint_{\mathbb{CP}^1} \frac{F}{2\pi}\right)}=1\,,
\end{eqnarray}
where $e$ is the $U(1)$ charge of the fermions and we used Gauss'  law. Then the minus sign that arises from parallel transporting the spinors can be cancelled by the minus sign  arising from propagating the $U(1)$ charges. Thus, one can consistently define charged spinors in this $U(1)$ background. This generalized spin structure is called a spin$^c$ structure. 

To obtain the quantization condition on the $U(1)$ flux,   we use $F=CK$, as discussed above, along with the expression of the K\"{a}hler $2$-form in (\ref{2-form in polar}). We take the limit $r\rightarrow \infty$ and integrate  eq.~(\ref{quantization1}) over the $\S^2$ parametrized by $\theta$ and $\phi$, recall (\ref{metricpolar}). We find $\oint_{\mathbb{CP}^1} K = - 2 \pi$, and obtain
\begin{equation}
1 = e^{i2\pi\left(\frac{1}{2}+ \frac{e C}{2 \pi} \oint_{\mathbb{CP}^1} K \right) } = e^{i 2 \pi \left({1 \over 2} - e C\right) }~.
\end{equation}
Thus, the quantization condition is $eC=m+\frac{1}{2}$ with $m\in \mathbb Z$. Without loss of generality we take $e=1$ and conclude that the necessary condition to define spinors on $\mathbb{CP}^2$ is to turn on the quantized monopole field
 \begin{eqnarray}
 F=\left(m+\frac{1}{2}\right)K\,.
 \end{eqnarray}

 As described in the main text, we also consider turning on the color, flavor, and baryon backgrounds (\ref{fluxes in Cartan}), reproduced here for convenience
 \begin{eqnarray}
F^{(c)}=\bm H^c\cdot \bm \nu^c m^c K\,,~
F^{(f)}=\bm H^f\cdot \bm \nu^f m^f K\,,~
F^{B}=\left(\frac{1}{2}+\frac{n^c}{N_c}m^c+\frac{n^f}{N_f}m^f\right)K\,.
\label{fluxes in Cartan1}
\end{eqnarray}
Notice that these are  embedded into the Cartan subalgebras of $SU(N_c)$ and $SU(N_f)$ and represent a generalization of the BCF 't Hooft flux backgrounds on $\mathbb T^4$ studied in \cite{Anber:2019nze}. 
When the $U(1)$ background $F = CK$ is replaced by (\ref{fluxes in Cartan1}), we obtain, instead of (\ref{sphere}),  for $\Psi$ of unit charge under baryon number, in a representation of $N_c$-ality $n_c$ and $N_f$-ality $n_f$,
\begin{eqnarray}
\label{sphere2}
\Psi(s=1)&=&e^{i2\pi \left({1\over 2} + \oint_{\mathbb{CP}^1} { F^B + n_f F^{(f)} + n_c F^{(c)} \over 2 \pi}  \right)}\Psi(s=0) \\
&=&e^{i 2 \pi \left({1 \over 2} - (\frac{1}{2}+\frac{n^c}{N_c}m^c+\frac{n^f}{N_f}m^f) - n_c \bm H^c\cdot \bm \nu^c m^c - n_f \bm H^f\cdot \bm \nu^f m^f   \right)} \Psi(s=0) \nonumber \\
&=&  \Psi(s=0)~,\nonumber
\end{eqnarray}
where the last equality follows from the fact that the fractional part of the eigenvalues of $\bm H^{c}\cdot \bm \nu^{c}$ is $- {1/N_{c}}$ (and similar for $c \rightarrow f$). Thus the background (\ref{fluxes in Cartan1}), or eq.~(\ref{fluxes in Cartan}) of the main text, is consistent with parallel transport on $\mathbb{CP}^1$.

The  Pontryagin number of the $U(1)$ bundle, using  $\int_{\mathbb{CP}^2} K\wedge K=\frac{8 \pi^2}{2}$, is given by
 \begin{eqnarray}
 {\cal P}=\frac{1}{8\pi^2}\int_{\mathbb{CP}^2} F\wedge F=\frac{1}{2}\left(m+\frac{1}{2}\right)^2\,, 
 \end{eqnarray}
which combines with (\ref{gravity index}) to give the full Dirac  index in the combined  $U(1)$ and $\mathbb{CP}^2$ background
\begin{eqnarray}
{\cal J}_D=\frac{1}{8\pi^2}\int_{\mathbb{CP}^2} F\wedge F+\frac{1}{192\pi^2}\int_{\mathbb{CP}^2}\mbox{tr}\left[R\wedge R\right]=\frac{m}{2}\left(m+1\right)\in \mathbb Z\,,
\end{eqnarray}
which now has integer values.\footnote{The zero modes of the Dirac operator on $\mathbb{CP}^2$ were studied and explicitly constructed in \cite{Kirchberg:2004za}.} 

Likewise, the Dirac index for the fermions of (\ref{sphere2}), in the background (\ref{fluxes in Cartan1}), is
\begin{eqnarray}
{\cal J}_D=T_{{\cal R}^c}\mbox{dim}_{{\cal R}^f}Q^c+ T_{{\cal R}^f}\mbox{dim}_{{\cal R}^c}Q^f+\mbox{dim}_{{\cal R}^f}\mbox{dim}_{{\cal R}^c}\left(Q^B+Q^G\right)\,,
\label{Dirac index2}
\end{eqnarray}
also given in (\ref{Dirac index}) of the main text, which is also an integer. Here, $Q^B=\frac{1}{8\pi^2}\int F^B\wedge F^B$ and $Q^{c/f}=\frac{1}{8\pi^2}\int\mbox{tr}\left[ F^{(c)/(f)}\wedge F^{(c)/(f)}\right]$,  explicitly given by
\begin{eqnarray}
\nonumber
Q^c=\frac{\left(m^c\right)^2}{2}\left(1-\frac{1}{N_c}\right)\,, ~Q^f=\frac{\left(m^f\right)^2}{2}\left(1-\frac{1}{N_f}\right)\,, ~
Q^{B}=\frac{1}{2}\left(\frac{1}{2}+\frac{n^c}{N_c}m^c+\frac{n^f}{N_f}m^f\right)^2\,.\\
\label{fractional top charges2}
\end{eqnarray}

Finally, we note that one can use equations (\ref{sphere2}, \ref{Dirac index2}, \ref{fractional top charges2}) to identify gauge theories that can be consistently formulated on  $\mathbb{CP}^2$ {\it without turning on global symmetry backgrounds,} i.e. by only modifying the conditions on the gauge bundles being summed over in the path integral. Constructions of this type were recently used to uncover a new $SU(2)$ anomaly \cite{Wang:2018qoy} on non-spin manifolds (note that in our examples
 all fermions can be given gauge invariant mass and there is no analogue of the new $SU(2)$ anomaly).

The simplest such case \cite{Bi:2018xvr}  is that of an $SU(2)$ theory with $N_f$ Dirac fundamental flavors. To see this from our equations,  take $N_c =2, m^c =1, n^c=1$, $Q^B = Q^f = 0$, and check that (\ref{sphere2}) holds and (\ref{Dirac index2}) is an integer (for any single flavor).  This $SU(2)$ QCD(F) with $N_f$ flavors was  interpreted in \cite{Bi:2018xvr} as  emerging near a  quantum critical point of a theory of only bosons (heuristically, this is because all gauge invariant operators are bosonic).

Other examples (involving both $SU(2)$ and other gauge groups) are  discussed in \cite{Wang:2018qoy,Davighi:2020bvi}. Within the class of theories  considered in this paper (specified in Section \ref{vectorlikedef}) the ones that do not require global symmetry backgrounds to be consistently formulated on $\mathbb{CP}^2$ must obey
 \begin{equation}
\label{condition1}
{1\over 2} +  {n_c m^c \over N_c} \in \Z~, ~~ T_{{\cal R}^c}{(m^c)^2 \over 2}(1 - {1 \over N_c}) - {1 \over8} \mbox{dim}_{{\cal R}^c}  \in \Z~ ,\end{equation} where   the second condition, the integrality of the index, should hold once the first is obeyed.
We have not exhaustively studied the solutions of the above conditions for general $n_c$, ${\cal R}_c$ and will only note a few simple cases. The first is QCD(F) with $N_f$ Dirac flavors and an $SU(N_c = 2k)$ gauge group. As in the $SU(2)$ theory of \cite{Bi:2018xvr}, it is easy to see that all gauge invariant operators are bosons (or that (\ref{condition1}) holds). The second set of theories where (\ref{condition1}) is easily seen to hold is  QCD(S/AS) with $N_f$ S/AS Dirac flavors and an $SU(N_c=4k)$ gauge group. As in the other examples, here also all gauge invariants (e.g. baryons and mesons) are bosons.

  \bibliography{References.bib}
  
  \bibliographystyle{JHEP}
  \end{document}